\documentclass[referee]{raa}            

\usepackage{graphicx,times}             
\usepackage{natbib}
\usepackage{amssymb,amsmath}
\bibpunct{(}{)}{;}{a}{}{,}

\usepackage[a4paper=true,dvipdfm=true,pagebackref=true]{hyperref}
\hypersetup{colorlinks = true, linkcolor = green, anchorcolor = red, citecolor = blue, filecolor = red, pagecolor = red, urlcolor = red}

\usepackage{boxedminipage}
\newenvironment{edesc}[1]{\begin{boxedminipage}[t]{#1}\begin{description}}{\end{description}\end{boxedminipage}}
\begin{document}

   \title{Stellar Spectra Classification and Feature evaluation Based on Random Forest
}

   \volnopage{Vol.0 (20xx) No.0, 000--000}      
   \setcounter{page}{1}          

   \author{Xiang-Ru Li
   \and Yang-Tao Lin
   \and Kai-Bin Qiu
   }

   \institute{School of Mathematical Sciences, South China Normal University,
             Guangzhou 510631, China; {\it xiangru.li@gmail.com}\\
\vs\no
   {\small Received~~2018 December 10; accepted~~2019~~March 18}}

\abstract{ With the availability of multi-object spectrometers and the designing \& running of some large scale sky surveys, we are obtaining massive spectra. Therefore, it becomes more and more important to deal with the massive spectral data efficiently and accurately. This work investigated the classification problem of stellar spectra under the assumption that there is no perfect absolute flux calibration, for example, the spectra from Guoshoujing Telescope (the Large Sky Area Multi-Object Fiber Spectroscopic Telescope, LAMOST). The proposed scheme consists of the following two procedures: Firstly, a spectrum is normalized based on a 17th polynomial fitting; Secondly, a random forest (RF) is utilized to classifying the stellar spectra. The experiments on four stellar spectral libraries show that RF has a good classification performance. This work also studied the spectral feature evaluation problem based on RF. The evaluation is helpful in understanding the results of the proposed stellar classification scheme and exploring its potential improvements in future.
}

   \authorrunning{X.-R. Li, Y.-T. Lin \& K. B. Qiu}            
   \titlerunning{Stellar Spectra Classification and Feature evaluation Based on Random Forest}  

   \maketitle

%
%
\section{Introduction}           
\label{sect:Intro}

With the development of modern telescopes, massive spectra have been and is being obtained.
In this massive spectrum scenario, traditional manual data processing methods and the
schemes with many human interventions can not meet the actual needs. Therefore, automatic
classification is an imperative problem in large sky surveys and attracts much
attentions \citep{AJ:Gulati:1994,MNRAS:Hippel:1994,Book:Gray:2009,MNRAS:Crowther:2011}.

Therefore, a series of schemes are investigated for automatic classification of spectra
in recent thirty years. Two most widely used schemes are template matching and artificial neural networks (ANN).
The template matching method is implemented by minimizing some metric distances or maximizing some kinds of similarity between a reference spectrum and a spectrum to be classified \citep{Kurtz:1984,MKP:LaSala:1994,NA:Malyuto:2002,MSAI:Giridhar:2006,AJ:Lee:2008,RAA:Duan;2009,AJ:Gray:2016}, for example , $\chi^2$ minimization. The ANN method classifies a spectrum by establishing a mapping from a spectrum to its spectral type or subtype \citep{PASP:Bailer-Jones:1997,MNRAS:Bailer:1998,MNRAS:Singh:1998,AJ:Weaver:2000,ChAJ:Bailing:2005,AS:Bazarghan:2008,BASI:Mahdi:2008,AA:Navarro:2012,ASS:Kheirdastan:2016}. Expert system \citep{AJ:Manteiga:2009,AJ:Gray:2014}, Support vector machine \citep{RAA:Liuchao:2015,ASS:Kheirdastan:2016} and K-means\citep{ISOP:Qin:2001,ASS:Kheirdastan:2016} are also investigated for the classification of stellar spectra.

In automatic classification of stellar spectra, a key problem is how to represent the information of a spectrum. This information representation problem is referred to as feature extraction in machine learning community. This information representation not only affects the accuracy of a spectral classification system, its robustness to noise \& calibration distortion, but also its interpretability/understandability. The interpretability means the difficulty to evaluate or track back the contribution of a specific wavelength range or spectral line in spectral classification. Good interpretability helps us understand our automatic classification scheme, its physical indications, and design an improved method by taking some physical knowledge into account.

Two typical information representation methods for a spectrum are spectral index \citep{MNRAS:Malyuto:1997,AJ:Lee:2008,AJ:Manteiga:2009,RAA:Liuchao:2015} and principal component analysis (PCA) \citep{ISOP:Qin:2001,BASI:Mahdi:2008,ASS:Kheirdastan:2016}. Spectral index can be an integration of spectral fluxes within a preset wavelength range, or some kinds of description of a spectral line, for example, full width half maximum (FWHM). The most superiority of spectral index is its interpretability. PCA is a kind of data compression method, and used to obtain a compact representation by statistically minimizing the difference between some spectra and their representations. Actually, the PCA representation is a linear sum of spectral fluxes. Therefore, one typical limitation of the PCA is difficult to evaluate/trace back the contribution of a local wavelength range of a spectrum, which is closely related with the interpretability of the computed results.

This work studies the automatic classification of stellar spectra using a random forrest. A random forrest consists of a series of decision trees. A decision tree makes spectral classification using a small subset of fluxes on automatic selected wavelength positions. This result that the random forrest have the potential superiority of interpretability. In application, furthermore, the effects of noise and calibration imperfects can vary from spectrum to spectrum, and from wavelength to wavelength. Fortunately, the choices of effective wavelength positions are different from tree to tree in random forrest.  This diversity on wavelength selection makes random forest achieve a good classification performance by adaptively selecting an appropriate combinations of wavelength positions in the competition between the decision trees of a random forest. Therefore, this work investigate the stellar spectral classification problem using the random forrest.

\section{Architecture of the proposed scheme and data preprocessing}\label{Sec:Scheme_Preprocess}

\begin{figure*}[!htp]
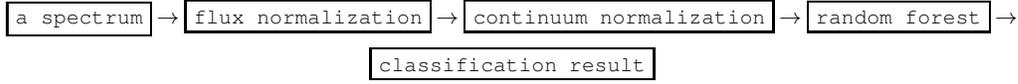

\centering
 \footnotesize{
\boxed{\texttt{a~spectrum}} $\rightarrow$ \boxed{\texttt{flux~normalization}} $\rightarrow$ \boxed{\texttt{continuum~normalization}} $\rightarrow$ \boxed{\texttt{random~forest}} $\rightarrow$ \boxed{\texttt{classification~result}}}
\normalsize
\caption{A flowchart of the proposed classification scheme of stellar spectrum. }
\label{flowchart}
\end{figure*}

To reduce some negative effects from the incompleteness of flux calibration, we first do some preprocesses on observed spectra. Then, the stellar spectra are classified using RF. A flowchart of the proposed scheme is presented in Fig. \ref{flowchart}.
\vspace{0.5cm}

\subsection{Flux normalization}\label{Sec:Scheme_Preprocess:Flux_Nor}
In spectral data, the observed radiant energies from some celestial bodies with the same spectral type may vary greatly in magnitude due to detector sensitivity, the brightness of the celestial body or the distance from the Earth. Some negative effects from magnitude uncertainty can be eliminated or reduced by normalizing the flux. Suppose $\bf{x}$ is a spectrum, denoted as: $\bf{x}=(x_1,x_2, ..., x_n)^T$, which is a vector in a $n$-dimension space. The spectral flux can be normalized using the following formula \citep{SSA:XU:2006}:

\begin{align}
\bf{y} = \frac{\bf{x}}{\sqrt{\sum_{i=1}^nx_i^2}}
\end{align}

\subsection{Continuum normalization based on polynomial fitting}\label{Sec:Scheme_Preprocess:Con_Nor}
This paper assumes that there is no perfect absolute flux calibration.
 Therefore, the classification algorithms cann't be used to classify spectra firstly, but rather, the continuum normalization is. In this paper, a 17-th order polynomial fitting method is used to approximate the continuum spectrum in a stellar spectrum. Then, the continuum components computed from the spectra are removed, leaving the spectral lines. Finally, a classification algorithm RF is utilized to the processed spectrum.

\begin{figure}[!htbp]
\begin{minipage}[t]{0.48\linewidth}
\centering
\includegraphics[height=6cm,width=7.5cm]{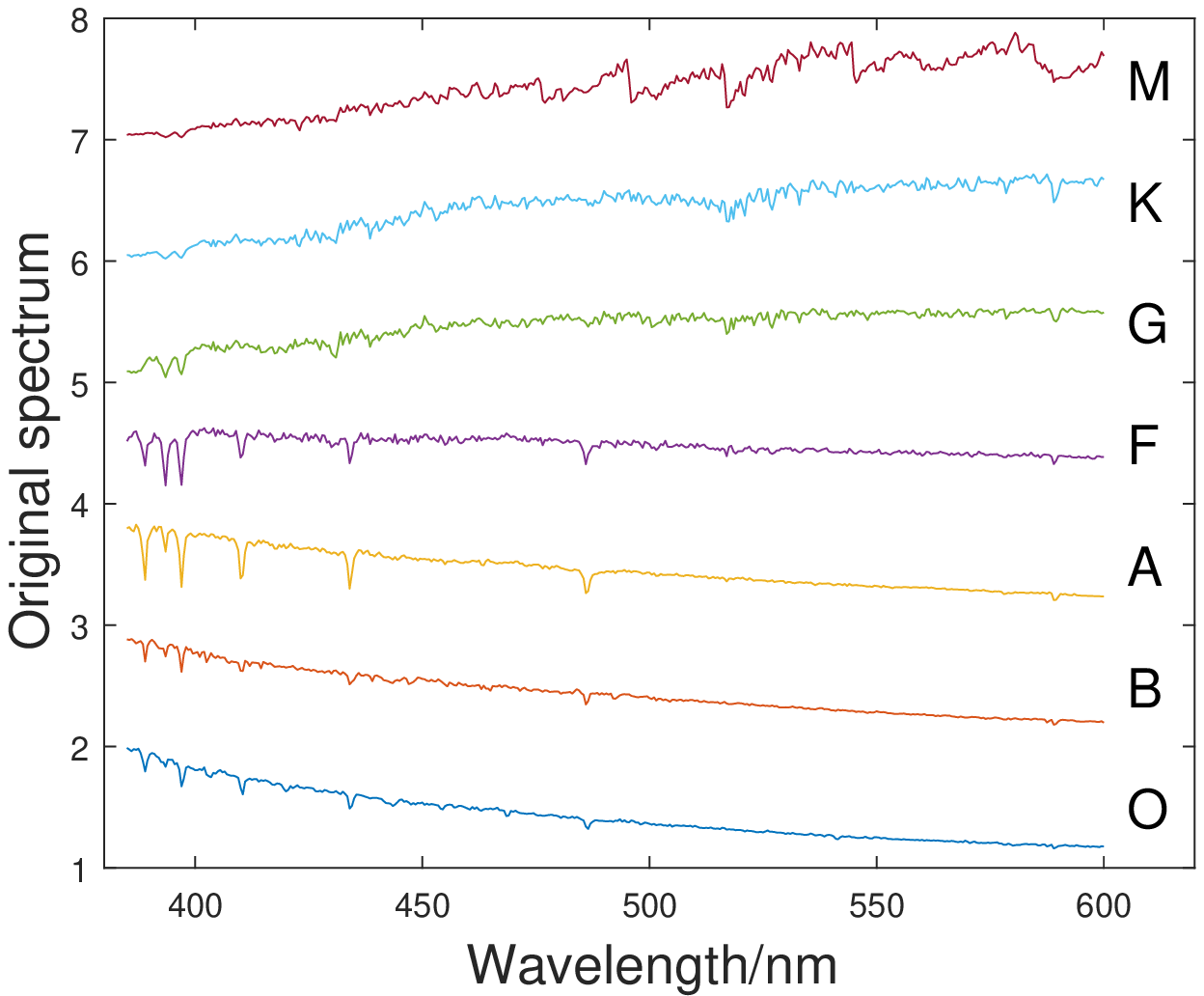}
\caption{Seven stellar spectra. nm: nanometer.}\label{Fig:original}
\end{minipage}
\hfill
\begin{minipage}[t]{0.48\linewidth}
\centering
\includegraphics[height=6cm,width=7.5cm]{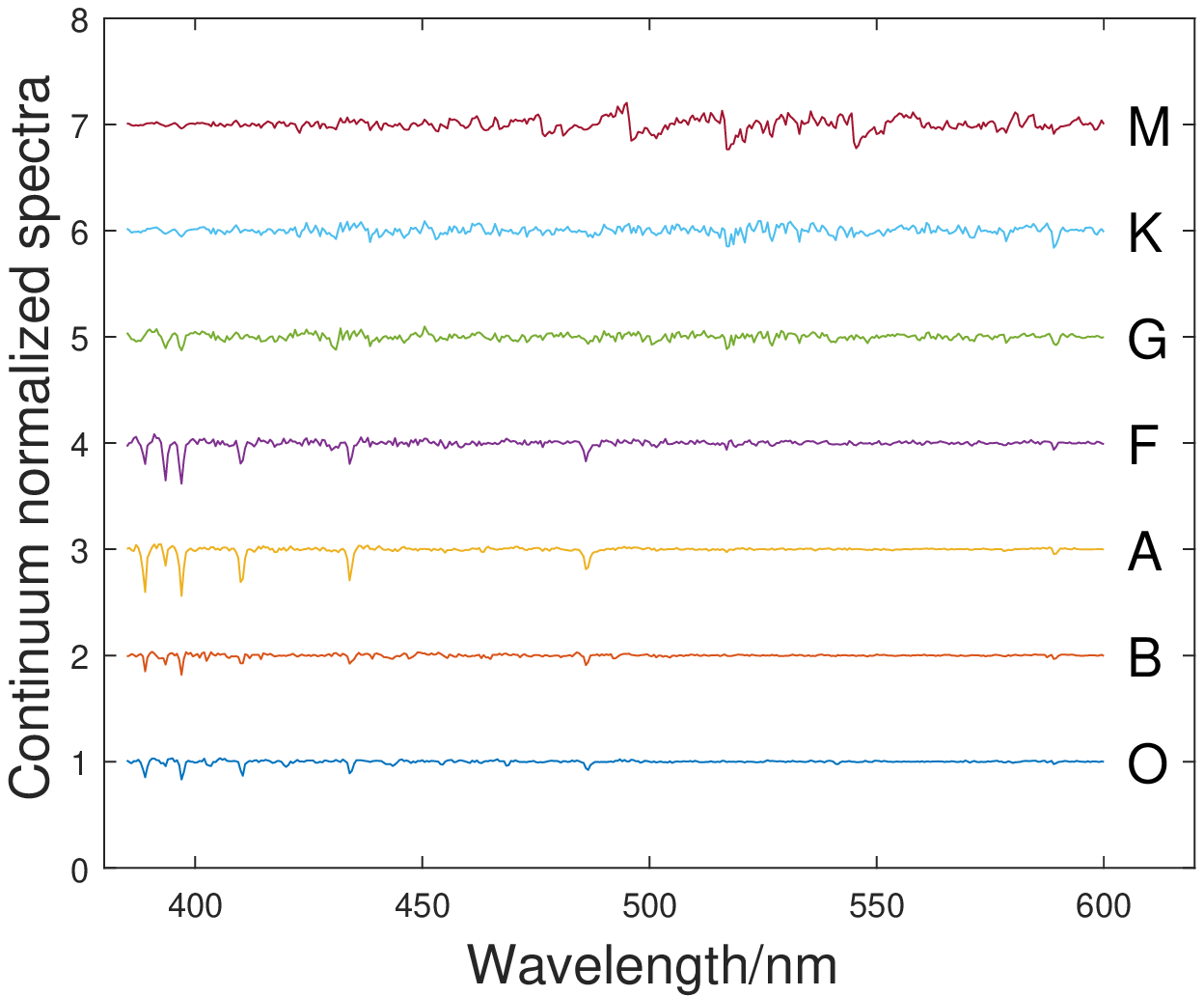}
\caption{The continuum normalized spectra in Fig. \ref{Fig:original}.}\label{Fig:continuum_normalized}
\end{minipage}
\end{figure}

Figures \ref{Fig:original} and \ref{Fig:continuum_normalized} show some continuum normalization results for some spectra from $O,B,A,F,G,K,M$ spectral types. Their continuums are fitted using a polynomial with order 17. The results show that the spectral line characteristics are reserved well.

\section{Classifying a stellar spectrum using a random forest}\label{Forest}\label{Sec:RF}
Random Forest (RF) algorithm is an extension of the traditional decision tree. RF is implemented by combining multiple decision trees. And a series of researches show that this combination improves classification performances evidently and increased the robustness to outliers and noise \citep{Journal:Ho:1998,Journal:Breiman:2001}. This section gives a brief introduction to the procedures building a RF.

Because a RF is established by assembling a series of decision trees, the decision tree will be introduced followed by the assembling scheme.

\subsection{Decision tree}\label{Sec:RF:Decision_Tree}

Decision tree classifier is a tree-like model.  In a decision tree, there are three types of nodes: a root node, some branch nodes, and some leaf nodes. If a node `S' accept signals from node `T', `T' is called the parent node of `S' and `S' is one child node of `T'. A root node don't have any parent node and there is a unique root node in a decision tree. Each leaf node has a parent node but doesn't have any child node. A branch node has one parent node and one or more child nodes. Signals can only move directly from a parent node to one of its child node.

A spectrum is classified by moving from root node to one leaf node. Suppose there are $K$ leaf nodes $\{\texttt{leaf}_i, i=1, \cdots, K\}$
and a training set $S$. Based on the leaf node that a training sample can reach, the training set can be split into $K$ subsets $S_1, S_2, \cdots, S_K$. A subset $S_k$ is labeled with the most frequent class in it, and denoted with $\texttt{label}_{k}$. If a spectrum to be dealt with moves from root to node $\texttt{leaf}_{k^*}$, this spectrum is classified into type $\texttt{label}_{k^*}$.

To construct a decision tree, one fundamental problem is to determine which data property should be used in one parent node. For interested readers, more introduction about decision tree can be found in \citep{Journal:Quinlan:1986,Book:Rokach:2008}.

\subsection{Random forest}\label{Sec:RF:RF}
Random forest does the classification of a stellar spectrum by establishing a series of decision trees and fusing their results. Suppose $S_{tr}$ is a training set consisting of $N$ spectra. A novel spectral set can be generated by randomly selecting $N$ samples from $S_{tr}$ with replacement, and can be referred to as a bootstrap set. `With replacement' means that in case of a spectrum being sampled into a bootstrap set, there is still probability that this spectrum is sampled in future for this bootstrap set. Therefore, some spectra in the training set may appear more than one times in a bootstrap set and some other spectra probably don't exist in this bootstrap set. By doing this, we can generate a number of bootstrap sets from a training set. From every bootstrap set, a decision tree is learned, a number of decision trees are learned from these bootstrap sets and form a random forest. More about the random forest can be found in \citet{Journal:Ho:1998,Journal:Breiman:2001,Book:hastie:2008}. An algorithm for building a random forest is as follows:

\begin{edesc}{0.95\textwidth}
\item[Algorithm 3.1] Random Forest Classifier
\item[Input] Training set $S_{tr}$, {test} set $S_{te}$, {number}, {$M$}, of trees
\item[Output] Estimated class label for every test sample
\item[Steps]
\item[1] Let $i =1$.
\item[2] Generate a bootstrap set from $S_{tr}$, and denote the set with $S_{tr}^{bs}$.
\item[3] Construct a decision tree from $S_{tr}^{bs}$, and denote this decision tree with $\texttt{tree}_i$. 
\item[4] Let $i = i+1$.
\item[5] Repeat the steps 2, 3 and 4 for M times.
\item[6] Estimate the class label for every test sample in $S_{te}$ using the $\{\texttt{tree}_i, i=1,\cdots,M\}$, and fuse the estimations from different decision trees by the majority votes as the final classification result.
\end{edesc}

\section{Experiments and discussions}
\subsection{Data sets}
The proposed scheme was evaluated on two sets of stellar spectral libraries. These two spectral sets are referred to as JSP data and LAMOST data. These data are described firstly in this subsection.

\subsubsection{The JSP spectral set}
This data set consists of 359 spectra from three representative stellar spectral libraries from \citet{AJS:Jacoby:1984}, \citet{AJS:Silva:1992}, and \citet{Catalog:Pickles:1998}. Each of the three libraries covers the spectral types from $O$ to $M$.

The Jacoby spectral library has 159 spectra with a configuration of $0.14nm$/pixel and a wavelength range of $351.1-742.8nm$. The Silva spectral library has 71 spectra with $0.5nm$/pixel and a wavelength range of $351.0-893.0nm$. The Pickles spectral library has 129 spectra with    $0.5nm$/pixel and a wavelength range of $360.0-900.0nm$. In order to analyze them on a same scale, all of the spectra are
resampled with a step $0.5nm$ using a linearly interpolation on the wavelength range of $385.0-600.0nm$.

\subsubsection{LAMOST Spectral Set}
From the DR5 data published by LAMOST, we select 6,000 stellar spectra with  SNRU, SNRG, SNRR, SNRI (the signal and noise ratio of u, g, r, i bands) are all higher than 20. To be consistent with the JSP spectral set,  all of the LAMOST spectra are also resampled with a step $0.5nm$ using a linearly interpolation on the wavelength range of $385.0-600.0nm$.

\subsubsection{spectral class representations}
In automatic classification of stellar spectrum, a fundamental problem is how to represent the class in a computer. We represent the spectral types
 (O, B, A, F, G, K, M) using the integers $1\sim 7$ respectively. Ten spectral subtypes are represented by
the product of the subtype number and 0.1. For example, the category of A0 star is
denoted as 3.0, and the category of star of F5 is as 4.5.

\subsection{Experimental results}

\begin{figure*}[!htbp]
\begin{minipage}[t]{0.49\linewidth}
\centering
\includegraphics[height=5.6cm]{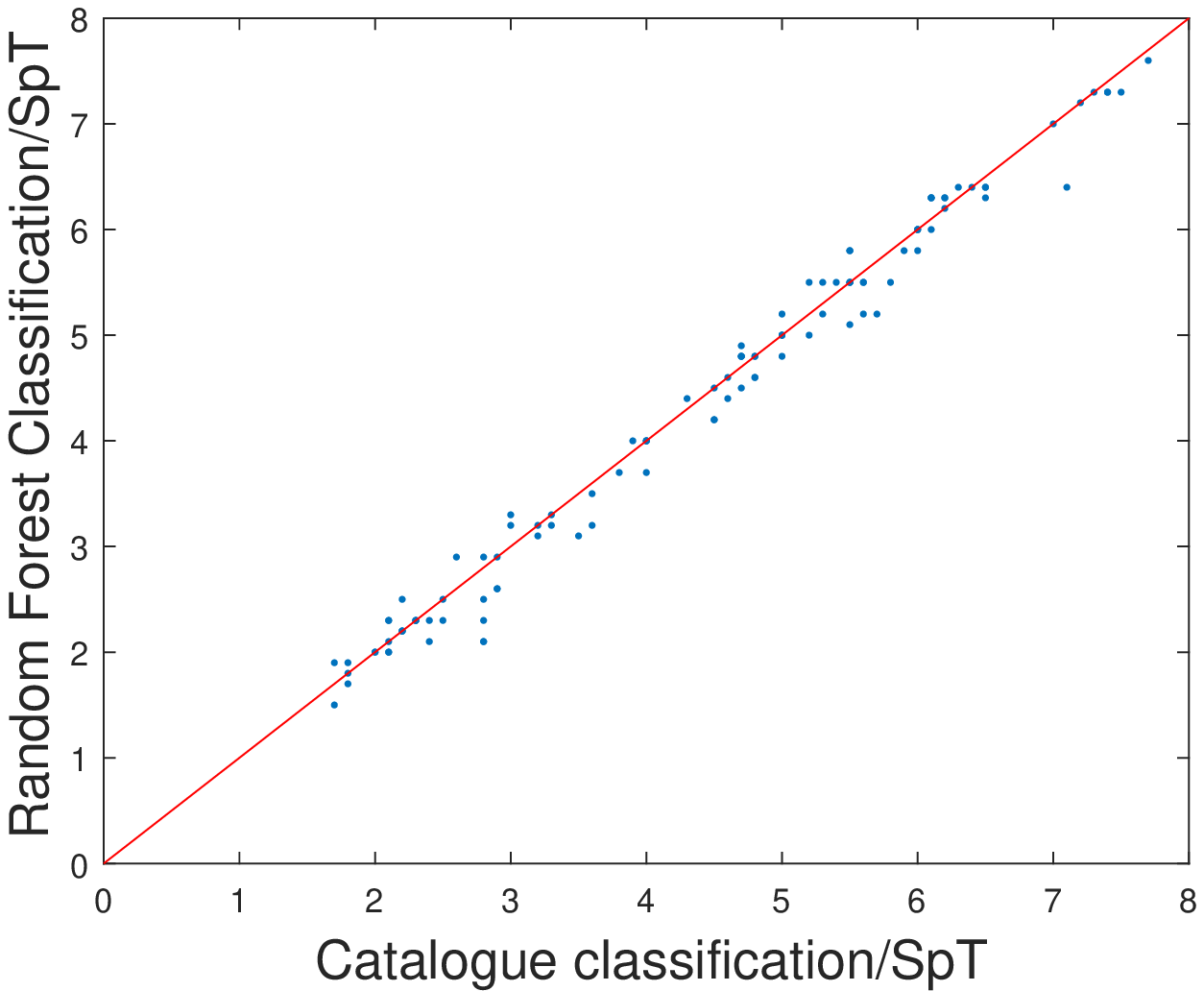}
\caption{Experimental results on JSP spectral set. The vertical axis is the estimation from the proposed scheme and the horizontal axis is the reference type. SpT: spectral types.}\label{classification_JSP}
\end{minipage}
\hfill
\begin{minipage}[t]{0.49\linewidth}
\centering
\includegraphics[height=5.6cm]{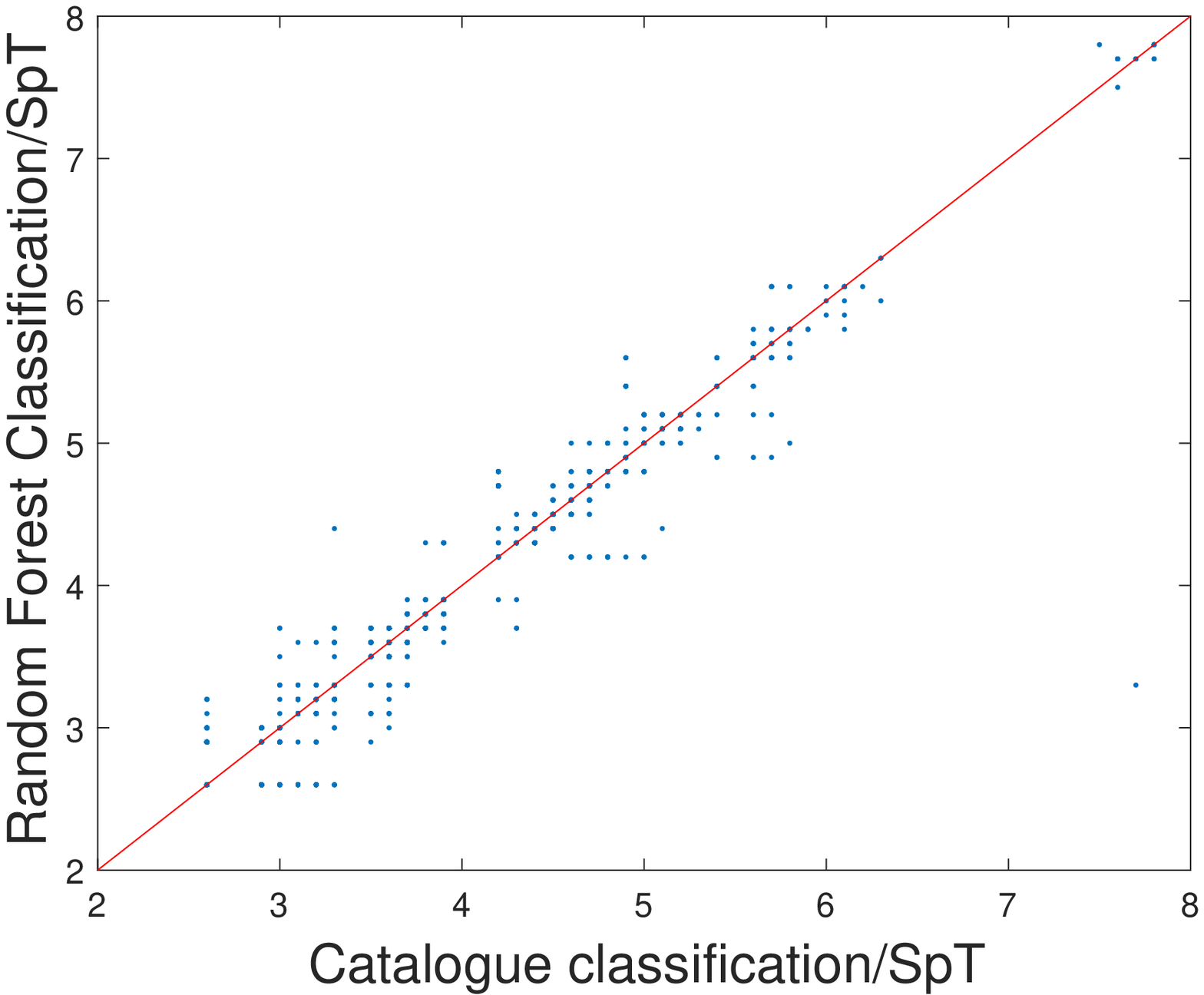}
\caption{Experimental results on LAMOST spectra. The vertical axis is the estimation from the proposed scheme and the horizontal axis is the reference type. SpT: spectral types.}\label{classification_LAMOST}
\end{minipage}
\end{figure*}

To evaluate the proposed scheme ( section \ref{Sec:Scheme_Preprocess} and section \ref{Forest}), the JSP spectral set is randomly divided into two subsets, a training set and a test set. The training set consist of $70\%$ of the JSP spectra and is used for learning the parameters of the random forest. The other $30\%$ spectra form the test set. The test set is used to evaluated the performance of the learned random forest. This evaluation result may depends on the dividing of training set and test set. To alleviate this issue and increase the objectiveness of the evaluation, we repeat the above-mentioned procedures 10 times, and take the average of ten experimental evaluations as the final result. The experimental results are presented in Figure \ref{classification_JSP}. The evaluation on LAMOST spectra set is conducted similarly and the results are presented in Figure \ref{classification_LAMOST}.

\begin{figure}[!htbp]
\centering
\includegraphics[width=6cm]{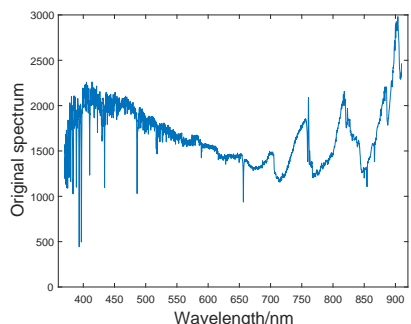}
		\setlength{\abovecaptionskip}{-2pt}
        \setlength{\belowcaptionskip}{-0.65cm}
\caption{The spectrum with the most spurious classification inconsistency in Fig. \ref{classification_LAMOST}. 
}
\label{type_M7_fig}
\end{figure}

The results in Fig. \ref{classification_JSP} show strong consistency between the estimation from the proposed scheme and the reference type. In experiments on LAMOST spectra (Fig. \ref{classification_LAMOST}), there exists a strong discrepancy on one spectrum. This discrepancy is indicated by the point on the right-down corner in Fig. \ref{classification_LAMOST}. This spectrum is presented in Fig. \ref{type_M7_fig}. Its reference type is M7 and the estimation is A3. After checking with helps from Dr. Xiao Kong in LAMOST, this is a spectrum of binary star with types M and A.

To quantitatively evaluate the performance of the proposed scheme, we use four measures --- mean of the squared difference ($\texttt{MSD}$), mean of the absolute difference ($\texttt{MAD}$), mean of difference ($\texttt{MD}$) and accuracy of spectral type ($\texttt{AST}$). Suppose $S=\{ (\bm{x}^i, y_i), i=1,\cdots,m \}$ is a set of spectra and their type labels; $\hat{y}_i$ is the estimation of $y_i$, where $m$ is an integer representing the number of spectra in $S$. On $S$, the MSD, MAD and MD are defined as follows:
\begin{equation}
   \texttt{MSD}(S)=\sqrt{\frac{\sum\limits_{i=1}^m(y^{(i)}-\hat{y}^{(i)})^2}{m}},
\end{equation}
\begin{equation}
   \texttt{MAD}(S)=\frac{\sum\limits_{i=1}^m|y^{(i)}-\hat{y}^{(i)}|}{m},
\end{equation}
\begin{equation}
   \texttt{MD}(S)=\frac{\sum\limits_{i=1}^m(y^{(i)}-\hat{y}^{(i)})}{m}.
\end{equation}

Suppose there are $n$ spectra in $S$ whose estimated spectral type are consist with their reference value, the AST is defined
\begin{equation}
\texttt{AST}(S) = n/m.
\end{equation}
Some quantitative evaluation results are presented in Table \ref{tab1}.

\begin{table}[!htbp]
\centering
\caption{Quantitative performance evaluation. FN: flux normalization, CN: continuum normalization.}\label{tab1}
\begin{tabular}{ccccc}
  \hline \hline
spectral set    &$\texttt{AST}$   &$\texttt{MSD}$   &$\texttt{MAD}$   &$\texttt{MD}$\\
 JPS spectra    &0.9537           &0.2151           &0.1481           &0.0574\\  \hline
LAMOST spectra  &  0.9377         &  0.1954         &  0.0787         &  0.0067\\
  \hline \hline
\end{tabular}
\end{table}

\subsection{Effects of spectrum preprocesses}
In the proposed scheme (Fig. \ref{flowchart}), two essential procedures are flux normalization and continuum normalization. There are more or less
deviations and distortions in observed spectrum. Therefore, these two preprocessing procedures evidently improved the spectral classification performance
on both JPS data and LAMOST data (Table \ref{tab1_normalization}). Particularly, the JPS spectra are observed with multiple telescopes and calibrated
using multiple pipelines, and more variety of calibration deviation and distortions exist on them. Therefore, much more performance improvement are
observed on JPS spectra than LAMOST spectra  (Table \ref{tab1_normalization}).

\begin{table}[!htbp]
\centering
\caption{Effects of spectrum preprocesses. IE: index of an experiment,
FN: flux normalization, CN: continuum normalization.}
\label{tab1_normalization}
\begin{tabular}{ccccccc}
  \hline \hline
 IE   & FN    &CN    &AST   &$\texttt{MSD}$   &$\texttt{MAD}$   &$\texttt{MD}$\\
 \hline
 \multicolumn{7}{c}{ (a) On JPS spectra} \\ \hline
1      & no    &  no  &0.4630    &1.9975           &1.2120           &1.0954\\
2      &yes    &{\small{no}}&0.8333&0.2287&0.1620&0.0250\\
3      &no&{\small{yes}}&0.4907&2.0000&1.2231&1.1472\\
4&yes&{\small{yes}}&0.9537&0.2151&0.1481&0.0574\\  \hline
\multicolumn{7}{c}{(b) On LAMOST spectra} \\ \hline
5&no&no&0.8391&0.3025&0.1642&0.0175\\
6&yes&{\small{no}}&0.9289&0.2242&0.0723&0.0067\\
7&no&{\small{yes}}&0.9091&0.2152&0.0999&0.0043\\
8&yes&{\small{yes}}&0.9377&0.1954&0.0787&0.0067\\
  \hline \hline
\end{tabular}
\end{table}

\subsection{Comparisons with related works in literature}
\citet{SSA:ZHANG:2009} studied the classification of stellar spectrum using a non-parametric regression method with a continuum spectrum normalization on the JSP data, and achieved an accuracy of  $MSD=0.3226, MAD=0.2554$. \citet{ASS:Kheirdastan:2016} obtained three accuracies of $MSD=1.39,1.53,1.64$ using an Artificial Neural Network, the SVM and K-means methods combined with PCA on some spectra from SEGUE-2 \citep{AJ:Yanny:2009} and SEGUE-1 of the SDSS III.
On the JSP spectra, \citet{SSA:LIU:2017} achieved an accuracy of  $MSD=0.2214, MAD=0.1632$ using the non-parametric regression method. The experimental results in Table \ref{tab1_normalization} show that random forest have good performance on both the JSP spectra and LAMOST spectra.

\section{Spectral feature evaluations}

The evaluation of variable effectiveness is a fundamental procedure to understand the potential physical interpretations and study more effective schemes. The random forest algorithm estimates the importance of a variable by looking at how much prediction error
increases in case of one variable being permuted with all others left unchanged. Conventional calculation methods of variable importance measure (VIM) in RF are divided into two types: One is based on the Gini index and the other is the Out-of-Bag (OOB) data error rate. The score statistics for the variable $X_j$ are denoted by $\text{VIM}_j^{(Gini)}$ and $\text{VIM}_j^{(OOB)}$ respectively. The interested readers are referred to \citet{Breiman:2002} for their definitions. In the existing literatures on RF, the $\text{VIM}_j^{(OOB)}$ score statistic is more extensive than the $\text{VIM}_j^{(Gini)}$ score statistic. There, this article ranks the importance of variables based on the $\text{VIM}_j^{(OOB)}$ score statistic.

\begin{figure*}[!htbp]
\begin{minipage}[t]{0.48\linewidth}
\centering
\includegraphics[height=6cm,width=7.5cm]{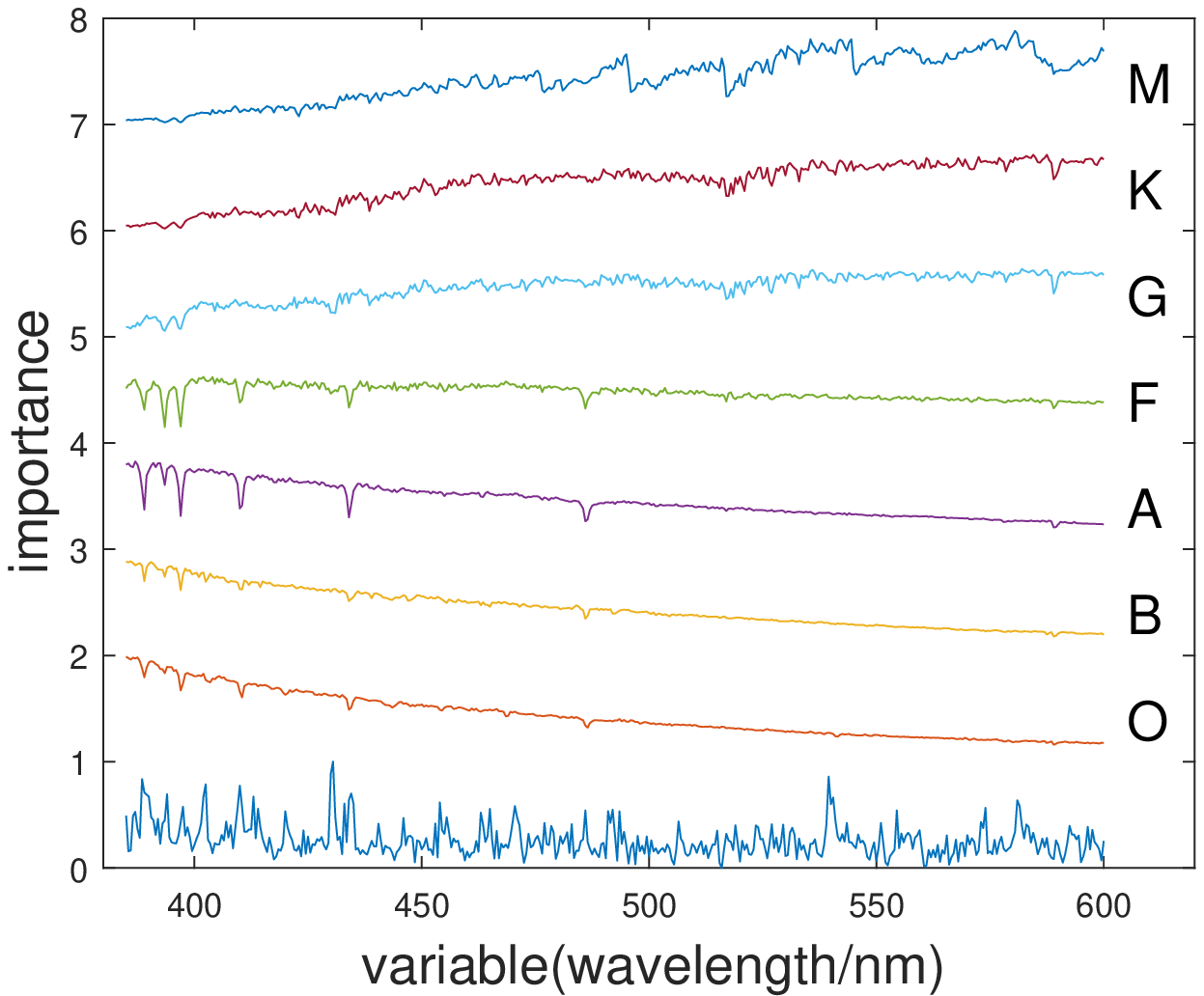}
\caption{The importance scores of stellar spectrum features on the JSP spectral data. The above seven curves are computed based on some spectra with spectral types $O\sim M$ respectively, the bottom curve indicates the effectiveness of the spectral fluxes. }\label{imp_JPS}
\end{minipage}
\hfill
\begin{minipage}[t]{0.48\linewidth}
\centering
\includegraphics[height=6cm,width=7.5cm]{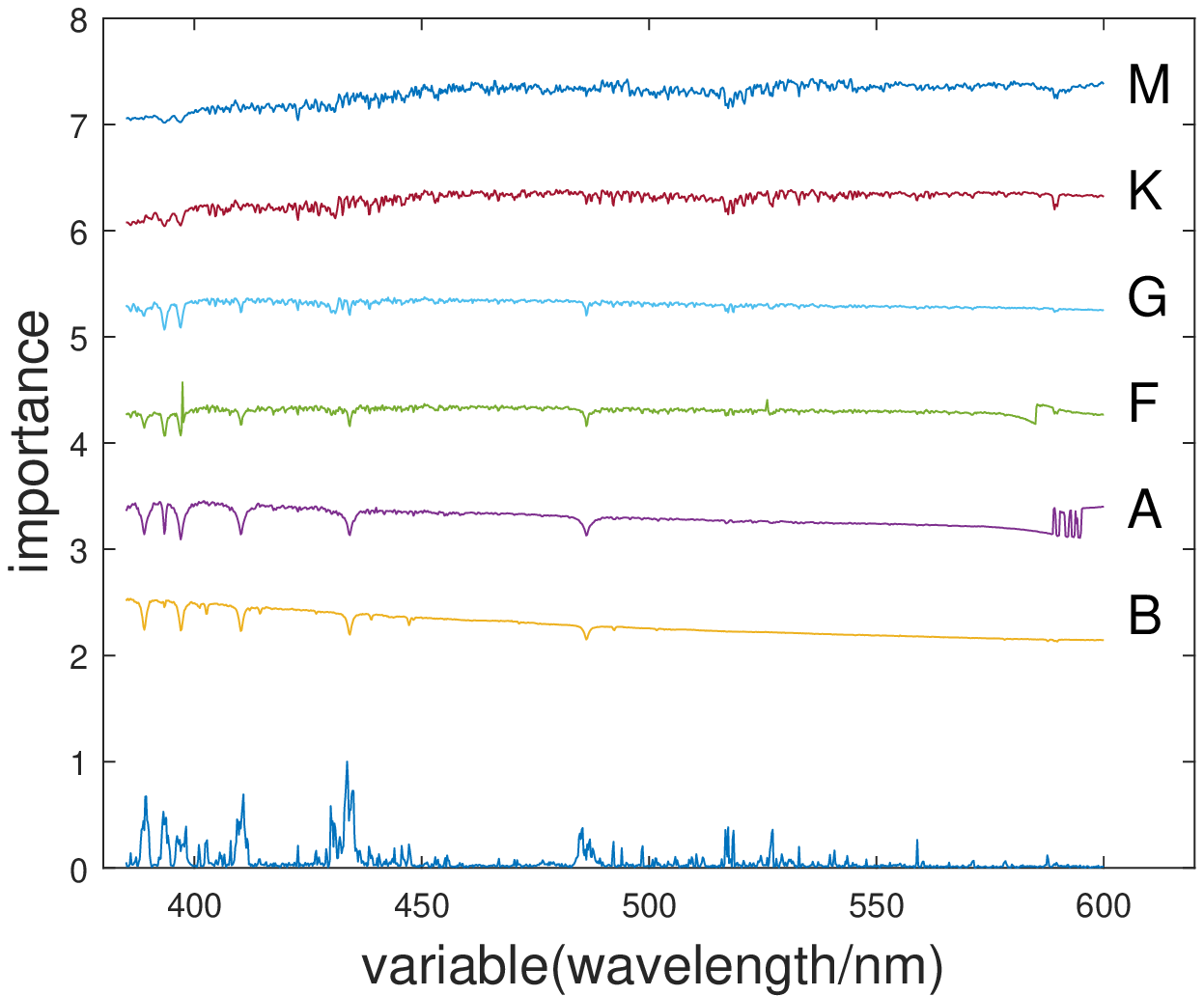}
\caption{The importance scores of stellar spectrum features on the LAMOST data. The above six curves are computed from spectra with spectral types $B\sim M$ respectively, the bottom curve indicates the effectiveness of the spectral fluxes. }\label{imp_LAMOST}
\end{minipage}
\end{figure*}

The evaluation results are presented in Figures \ref{imp_JPS} and \ref{imp_LAMOST}. In the figure \ref{imp_JPS}, the eight curves  from the bottom to the top are the importance scores of the spectral features at every wavelength computed from the JSP spectral data with spectral types from $O$ to $M$ respectively. The seven curves in Figure \ref{imp_LAMOST} from the bottom to the top are the importance scores of the spectral features at every wavelength computed from the LAMOST spectral data with spectral types from $B$ to $M$ respectively, showing the relationship between the important spectral variables and spectral lines of each type. The results from these figures show that for the spectral data from different systems, the important features selected by the Random Forest are approximately similar, which indicates that the random forest selects the important spectral lines of each type of spectrum as the basis for classification. The evaluation is helpful in understanding the results of the proposed stellar classification scheme and exploring its potential improvements in future.

\section{Conclusion}

Although there are a series of researches in literature for the automatic classification of stellar spectrum. However, the performance of automatic classification is still under improvementsㄛ especially for some spectra without flux calibration or only with relative flux calibration, for example, the spectra from LAMOST.

This work proposed a stellar spectrum classification scheme based on random forest and experimental results show its superiority on real spectral data. The characteristics of this work are the comprehensive investigation of the effects from flux normalization and continuum normalization. This work also studied the evaluation of spectral features. This evaluation is helpful in understanding the potential physical interpretations and designing more effective schemes.

\vspace{0.4cm}

\noindent This work is supported by the National Natural Science Foundation of
	China (grant No: 61273248, 61075033), the Natural Science
	Foundation of Guangdong Province (2014A030313425,
	S2011010003348), China Scholarship Council (201706755006), and the
Joint Research Fund in Astronomy (U1531242) under cooperative agreement between the National
Natural Science Foundation of China (NSFC) and Chinese Academy of Sciences (CAS).

\label{lastpage}

\end{document}